\documentclass[showpacs,twocolumn,amsmath,prb]{revtex4-1}
\usepackage{epsfig}
\usepackage[utf8]{inputenc}

\begin{document}
\title{Analysis of power-law exponents by maximum-likelihood maps}
\author{Jordi Bar\'o}
\email{jordibaro@ecm.ub.es}
\affiliation{Departament d'Estructura i  Constituents de la Mat\`eria.
Facultat  de F\'\i  sica.   Universitat de  Barcelona. Diagonal,  647,
E-08028 Barcelona, Catalonia.}
\author{Eduard Vives}
\email{eduard@ecm.ub.es}
\affiliation{Departament d'Estructura i  Constituents de la Mat\`eria.
Facultat  de F\'\i  sica.   Universitat de  Barcelona. Diagonal,  647,
E-08028 Barcelona, Catalonia.}
\begin{abstract}
Maximum-likelihood exponent  maps have been studied as  a technique to
increase the understanding and  improve the fit of power-law exponents
to experimental  and numerical  simulation data, especially  when they
exhibit both  upper and  lower cut-offs. The  use of the  technique is
tested  by analysing  seismological data,  acoustic emission  data and
avalanches  in  numerical simulations  of  the  3D-Random Field  Ising
model.  In  the  different  examples  we discuss  the  nature  of  the
deviations observed in the exponent maps and some relevant conclusions
are drawn for the physics behind each phenomenon.
\end{abstract}

\maketitle

\section{Introduction}

For the last few decades  the study of critical phenomena has received
a  great  deal  of  attention  in  many  different  areas  of  physics
\cite{Sornette2000}.  Criticality is  often identified by the presence
of statistical scale-free distributions of different magnitudes when a
system evolves in time or when it is driven by an external force.  The
physical  nature  of the  measured  response  can  be very  different:
energy,    displacement,    magnetization,    volume,    polarization,
resistivity, etc.  In all cases, when the sudden changes (often called
avalanches)  of  such magnitudes  exhibit  a statistical  distribution
compatible with a power law probability density function $g(x) dx \sim
x^{-\alpha}  dx$   one  gains   confidence  about  the   existence  of
criticality.  In some  cases criticality has also been  referred to as
``crackling noise'' \cite{Sethna2001}.  At criticality the response of
the system is characterized solely by the critical exponents $\alpha$.
The theoretical  understanding provided by the  use of Renormalization
Group techniques\cite{Binney1992} states, in many cases, that critical
exponents show a certain degree  of universality.  Thus, it has become
extremely  important to  determine the  values of  $\alpha$ to  a high
degree of accuracy and confidence.

Within  this framework,  it  is  important to  develop  tools to  test
power-law  behaviour of data  samples and  to fit  critical exponents.
The  development of  these statistical  tools cannot  be  done without
taking  into account  the  limitations inherent  to data  acquisition.
Typically  in   physics,  data  comes  from   experiment  or  computer
simulation  and in  both  cases  one has  boundaries  to the  proposed
scale-free behaviour.   These boundaries are not  necessarily sharp or
well defined.   In experiments  one finds unavoidable  noise deforming
the  power-law distribution  in the  small-event region  and different
kinds of instrument saturation  in the large-event region.  One should
take into  account the fact that  it is difficult  to find instruments
(amplifiers, voltmeters,  etc.)  that allow measurements  with a range
of more  than 5  decades.  In simulations  one also  finds unavoidable
limitations:  for instance,  one has  a minimum  lattice  parameter or
particle size that alters the small-event distribution and finite-size
effects deforming the large  events.  Since simulations with more than
$10^5$ particles  are scarce, it  is also difficult to  find power-law
distributions extending many decades  in numerical works. In addition,
the  existence  of deformations  in  the  region  of small  events  is
understood  not only  because of  the reasons  discussed above  with a
physical origin, but also because of a mathematical constraint: a pure
power-law  probability density  with $\alpha>1$  cannot  be normalized
without a theoretical lower limit $x_{min}$.

It  is  easy to  understand  that  a  naked-eye analysis  of  standard
histograms can  be easily fooled, not  only by the  lack of statistics
(insufficient data in the recorded  sample), but also by the anomalies
in  the large-  and small-event  regions.   The same  may happen  with
traditional  fitting methods  such as  the least-squares  method (both
linear  and non-linear) which,  for instance,  depends on  the binning
process that is performed in order to plot the histograms.

Many years  ago most of  the community adopted  the maximum-likelihood
(ML) estimation  method\cite{Goldstein2004, Bauke2007, Clauset2009} as
the safest  way to treat  data, although it  is still frequent  to see
papers  using alternative, error-prone  fitting methods.   Within this
scenario, the  work done  by M.E.J.  Newman  and co-workers  should be
pointed  out \cite{Clauset2009}.  Using  ML methods  they have  nicely
illustrated how  to test  the power-law character  of data and  how to
obtain good  estimations (and error bars) of  critical exponents.  One
of  the proposed  techniques consists  of studying  how robust  the ML
exponent is when the analysed  data is restricted to being higher than
an imposed lower cut-off $X_{low}$  that is varied by several decades.
By this method one studies  the deformation of the fitted exponent due
to undesired effects in the region of small events.

In this  paper we will  study the extension  of this technique  to the
analysis of  the ML exponent  as a function  of both an  imposed lower
cut-off $X_{low}$ and an imposed higher cut-off $X_{high}$.  This will
render  the so-called  ML  exponent maps  \cite{Peters2010}.  By  this
method we expect to be able to improve exponent estimation in the case
in which experimental or simulation data presents distortions not only
in the region of small events, but also in the large-event region.

In  section \ref{ML}  we  will revisit  the  ML method  and define  ML
exponent  maps. We  will include  a discussion  on  numerical methods,
evaluation of  error bars and  analysis of synthetic data  obtained by
pseudo-random number generation.  In  section \ref{Seis} we will apply
the proposed  analysis technique to  the study of  three seismological
catalogues  from Japan,  the San  Andreas  fault and  the very  recent
activity  in the  island of  El Hierro  (Canary Islands).   In section
\ref{Exp} we will study  three experimental cases corresponding to the
measurement  of  the  energy  of  acoustic  emission  (AE)  events  in
different phenomena in solids, from previous literature The three sets
of experiments have been carried  out with the same experimental setup
but  with different  experimental  constraints (noise,  amplification,
etc..).  The three cases correspond to (i) the compression of a porous
material (Vycor) \cite{Salje2011},  (ii) a cubic-tetragonal structural
transition  in  FePd \cite{Bonnot2008}  and  (iii) a  cubic-monoclinic
structural    transition   in    a   CuZnAl    shape    memory   alloy
\cite{Gallardo2010}.   In  Sec.~\ref{Sim} we  will  illustrate how  to
apply the  technique to the study  of numerical simulations  of the 3D
Random Field Ising model with  metastable dynamics which is one of the
prototypical      frameworks       for      avalanche      criticality
\cite{Sethna1993,Perkovic1995}.   Finally, in  Sec.~\ref{Con}  we will
supply a summary and draw conclusions.

\section{Maximum-likelihood exponent map}
\label{ML}

Let  us  consider  a  sample  of  measurements  $\lbrace  \mathrm{X}_i
\rbrace$  ($i=1,  \cdots,  N$)  that  we assume  to  be  statistically
independent. We will denote by capital letters $X_{max}$ and $X_{min}$
the smallest and  the largest values in the sample set.  Our aim is to
model this set with a power-law probability density:
\begin{equation}
g(x)dx=\frac{x^{-\gamma}}{\zeta (\gamma)} dx \;\quad \; x_{min}<x<x_{max}
\end{equation}
where  $\gamma$  is  the  exponent   that  we  wish  to  estimate  and
$\zeta(\gamma)$ denotes a  normalization function.  This normalization
function will depend on $\gamma$ and the theoretical upper ($x_{max}$)
and lower ($x_{min}$)  bounds, as will be discussed  below.  (The case
$x_{max} \rightarrow  \infty$ is a  particular case of  our analysis).
Note that these theoretical  limits will not necessarily coincide with
the   maximum  $X_{max}$   and   minimum  $X_{min}$   values  in   the
sample. Nevertheless, given the power-law character of the probability
density we expect that $X_{min}$ will be much closer to $x_{min}$ than
$X_{max}$  to  $x_{max}$.   Table  \ref{TAB1}  clarifies  the  generic
definitions of the  limits and cut-offs that will  be used within this
section.
\begin{table}
\begin{tabular} {|c|l|}
\hline symbols & meaning  \\ \hline 
$x_{min}$, $x_{max}$ & sharp bounds of the density $g(x)$  \\ 
$X_{min}$, $X_{max}$ & extreme values in the sample $\{X_i\}$ \\ 
$X_{low}$, $X_{high}$ & cut-offs imposed on the sample for the analysis \\ 
\hline
\end{tabular}
\caption{\label{TAB1}  Terminology used  in  this work  to define  the
generic bounds and cut-offs used  for the theoretical analysis in this
section}
\end{table}
The Likelihood  function $\mathcal{L}$  is defined as  the probability
that  the set of  measurements $\lbrace  \mathrm{X}_i \rbrace$  can be
obtained by the proposed model:
\begin{equation}
\mathcal{L}(\gamma ) = \prod_{i=1}^N g(\mathrm{X}_i )
\end{equation}
The Maximum Likelihood (ML) estimation method consists of choosing the
value  of $\gamma$ that  maximizes the  Likelihood Function,  i.e. the
value that makes the sample that  we have obtained the most likely one
to have occurred.

In  order to  evaluate  the deviations  of  data with  respect to  the
proposed  model,  in this  work  we  will  perform ML  estimations  by
restricting  the  original  data  within  the  imposed  lower  cut-off
$X_{low}$ and  the imposed  higher cut-off $X_{high}$,  different from
the  theoretical limits  $x_{min}$ and  $x_{max}$ which  are generally
unknown.   We will use  the symbol  $\hat{\gamma}$ to  distinguish the
exponent  estimated within  a  restricted interval  from the  exponent
estimated from the whole available  sample. We will use $n$ ($n<N$) to
denote the  number of  data of the  restricted set.  The normalization
factor of the probability density with imposed cut-offs is:
\begin{equation}
\zeta (  \hat{\gamma}) = \int_{X_{low}}^{X_{high}}  x^{- \hat{\gamma}}
dx            =           \frac{X_{low}^{1-{\hat{\gamma}}}           -
X_{high}^{1-{\hat{\gamma}}}}{\hat{\gamma} -1 }
\end{equation}
The  best  estimation  of  $\hat{\gamma}$  is  consequently  found  by
maximizing the likelihood function:
\begin{equation}
0    =    \frac{\partial    \ln\mathcal{L}(\hat{\gamma})}    {\partial
\hat{\gamma}}=-\sum_{\lbrace  X_{low} <  X_i  < X_{high}  \rbrace}^{n}
\ln(\mathrm{X_i})   -  n   \dfrac{\zeta   '(  \hat{\gamma})}   {\zeta(
\hat{\gamma})}
\end{equation}
where:
\begin{equation} 
\dfrac{\zeta     '(     \hat{\gamma})}{\zeta(     \hat{\gamma})}     =
\dfrac{1}{1-\hat{\gamma}}    -   \dfrac{X_{high}^{1-\hat{\gamma}}   \ln
X_{high}            -           X_{low}^{1-\hat{\gamma}}           \ln
X_{low}}{X_{high}^{1-\hat{\gamma}}- X_{low}^{1-\hat{\gamma}}}
\label{cont}
\end{equation}
We should  mention that for  the case in  which data is  discrete (for
instance, in  many simulations of lattice models)  the above treatment
should be slightly  modified.  In this case, the  data consists of the
frequencies of occurrence $f(k)$ of  a discrete set of values $\lbrace
k \rbrace$ (which we will assume to be integers). We would like to fit
it   with   a  power-law   probability   function   (called  Zeta   or
Zipf)\cite{Goldstein2004}:
\begin{equation}
p(k)  = \frac{k^{-\gamma}}{\zeta(\gamma)}  \; \;  k_{min} \leq  k \leq
k_{max}
\end{equation}
Following the same  procedure as above, when we  restrict ourselves to
data   within   imposed  cut-offs   $K_{low}$   and  $K_{high}$,   the
normalization function is:
\begin{equation}
\zeta(\hat{\gamma})= \sum_{k=K_{low}}^{K_{high}} k^{-\hat{\gamma}}
\end{equation}
and the derivative of the Likelihood Function will read:
\begin{equation}
\frac{\partial         \ln\mathcal{L}}{\partial         \hat{\gamma}}=
-\sum_{\mathrm{k}=K_{low}}^{K_{high}}{f(\mathrm{k})\ln(\mathrm{k}) + N
  \dfrac{\sum_{K_{low}}^{K_{high}}         {\mathrm{k}}^{-\hat{\gamma}}
    \ln(\mathrm{k})}{\sum_{K_{low}}^{K_{high}}\mathrm{k}^{-\hat{\gamma}}}}
\label{disc}
\end{equation}
As opposed to  what happens in the case in which  one considers only a
lower  cut-off,  equations  (\ref{cont})  and (\ref{disc})  cannot  be
solved  analytically.  Thus,  in  this  work, we  will  use the  false
position method in order  to find roots \cite{Press1992}.  This method
generates  a sequence  of  recursively smaller  intervals that  always
include the root  of the equation.  The monotony  of the derivative of
$\ln\mathcal{L}$\cite{Miller2001}  ensures  that  the  false  position
method  always  converges  to  the  root.  We  have  chosen  arbitrary
starting values of $\hat{\gamma}_1=  1$ and $\hat{\gamma}_2= 3.5$, and
have   iterated   the    algorithm   $M$-times   until   an   interval
$(\hat{\gamma}_{M-1},\hat{\gamma}_{M})$  is  reached  with a  distance
smaller than $0.005$.

\begin{figure}[htb]
\begin{center}
\epsfig{file=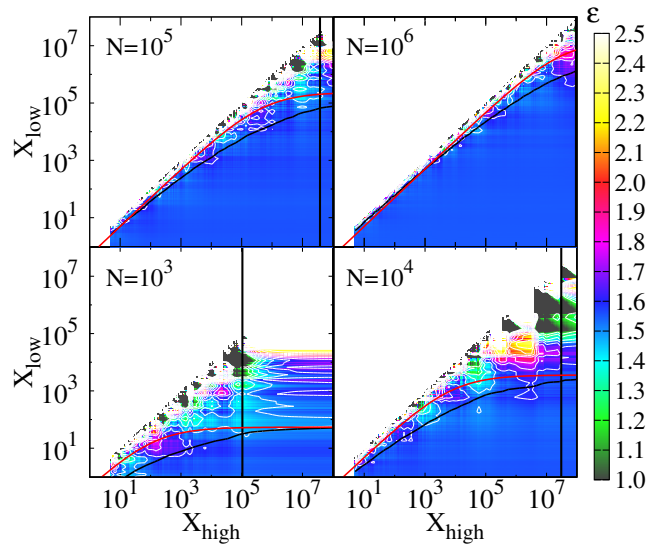,width=8.6cm,clip=}
\end{center}
\caption{\label{FIG1} Exponent maps  obtained from synthetic data. The
data  sets correspond  to a  theoretical probability  density function
with    exponent   $\gamma=1.55$    and    bounds   $x_{min}=1$    and
$x_{max}=10^{8}$.  The label $N$ indicates the size of the sample set.
The two  curved continuous lines  indicate the limits above  which the
standard  deviation becomes greater  than 0.05.   The red  curved line
corresponds to  the estimation using  Eq.~(\ref{errorclauset}) and the
black curved  line to Eq.~(\ref{error}).  The  vertical black straight
line marks the highest value in the sample $X_{max}$}
\end{figure}

By  changing the  $X_{low}$ and  $X_{high}$ cut-offs  we can  plot the
values of the  ML estimations of $\tilde{\gamma}$ using  a color scale
and   thus  obtain   the  exponent   map.   Examples   are   shown  in
Fig.~\ref{FIG1}  and throughout  the paper.  Contour lines  (in white)
will also  be shown separating exponent  values in steps  of 0.1.  The
maps exhibit  a triangular shape  since they are obviously  limited by
the  condition $X_{high}>X_{low}$.   The main  goal of  the map  is to
check the existence of a flat plateau (with an homogeneous color which
is free of contour lines) in  which the exponent is independent of the
cut-offs and thus  confirm the scale-free behaviour of  the data.  The
map also  allows anomalies  to be identified  that can  have different
origins, as will be discussed in  the following sections by the use of
examples.

One  of the advantages  of using  the false  position method  for root
finding  is that  we  straightforwardly obtain  an  estimation of  the
second derivative of $\ln \mathcal{L}$ at the maximum:
\begin{equation}
\left  .  \frac{\partial^2  \ln  \mathcal{L}}{\partial \hat{\gamma}^2}
\right    |_{max}\sim    \frac{    \left   .     \frac{\partial    \ln
\mathcal{L}}{\partial   \hat{\gamma}}    \right   |_M   -    \left   .
\frac{\partial \ln  \mathcal{L}}{\partial \hat{\gamma}} \right |_{M-1}
}{\hat{\gamma}_{M}-\hat{\gamma}_{M-1}}
\end{equation}
Assuming  Gaussian  behaviour of  the  Likelihood  Function (which  is
ensured by  the Central  Limit Theorem when  $n$ is large  enough) and
under  very  general  conditions\cite{Eadie},  the  second  derivative
provides us  with an  approximation to the  standard deviation  of the
estimated exponent.
\begin{equation}
\sigma_{\hat{\gamma}}=\left[{-    \left   .     \frac{\partial^2   \ln
\mathcal{L}}{\partial \hat{\gamma}^2} \right |_{max}}\right]^{-1/2}
\label{error}
\end{equation}
It is  easy to check  that this expression, when  $x_{max} \rightarrow
\infty$,     reduces      to     the     equation      proposed     in
Ref.~\onlinecite{Clauset2009}:
\begin{equation}
\sigma_{\hat{\gamma}} = \frac{\hat{\gamma} -1}{\sqrt{n}}
\label{errorclauset}
\end{equation}
%

\begin{figure}[htb]
\begin{center}
\epsfig{file=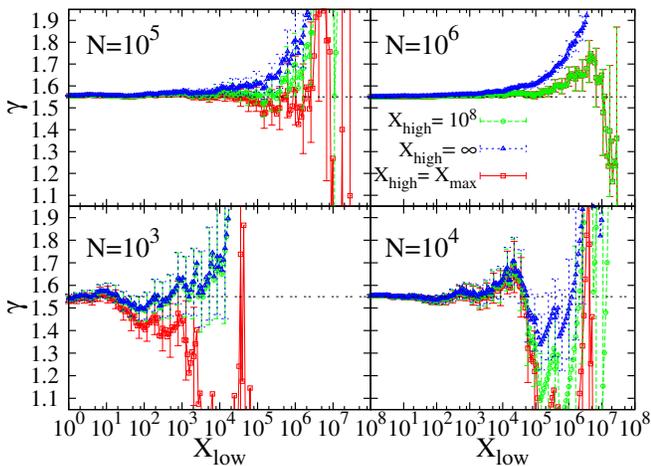,angle=270,width=8.6cm,clip=}
\end{center}
\caption{\label{FIG2}  Comparison  of the  behaviour  of the  exponent
fitted to synthetic data as  a function of the lower cut-off $X_{low}$
when    $X_{high}$   is    fixed.    Red    squares    correspond   to
$X_{high}=x_{max}=10^8$ (the theoretical limit of the synthetic data),
green circles  to $X_{high}=X_{max}$ (the  maximum value found  in the
sample) and blue triangles to $X_{high}=\infty$.}
\end{figure}

As a first test of the  usefulness of the ML exponent maps, we studied
synthetic data  \cite{Clauset2009} generated according  to a power-law
probability  density  with  theoretical   limits  $x_{min}  \leq  x  <
x_{max}$:
\begin{equation}
g(x)   dx  =   (\gamma  -1)   \frac{x^{-\gamma}}{x_{min}^{1-\gamma}  -
x_{max}^{1-\gamma}} dx
\label{syntetic}
\end{equation}
with  $\gamma=1.55$, $x_{min}= 1$  and $x_{max}=10^8$.   The synthetic
data samples were obtained using the RANECU generator \cite{James1990}
for  uniform random numbers  in the  $[0,1)$ interval  and transformed
using the method based on the inverse cumulative distribution function
$F^{-1}(z)$.   For the  proposed probability  density (\ref{syntetic})
the transformation  function that converts uniform  random numbers $z$
into the desired ones is given by:
\begin{equation}
x        =        F^{-1}(z)=\left[{x_{max}^{1-\gamma}       -        z
(x_{max}^{1-\gamma}-x_{min}^{1-\gamma})}\right]^{\frac{1}{1-\gamma}}
\end{equation}
Fig.~\ref{FIG1} shows the resulting maps corresponding to four samples
of  increasing sizes  $N$  as  indicated by  the  legends.  The  first
observation is that the size $N$  of the sample is crucial in order to
obtain a  clean plateau corresponding  to the correct  exponent.  Only
when   $N  \gtrsim  10^4$   does  the   plateau  extend   for  several
``square''-decades.

We have indicated the position of the maximum value $X_{max}$ obtained
in the sample  by a vertical black line. The  variations of the fitted
exponent observed from this line to the right are simply a consequence
of changing  the imposed upper  cutoff $X_{high}$ in a  region without
available  data due  to  the lack  of  statistics. This  lack of  data
between $X_{max}$  and $X_{high}$ is  relevant for the  fitting method
and, a priori, should be  taken into account. Furthermore, the imposed
cutoff $X_{high}$  can be moved above the  theoretical limit $x_{max}$
or set  to be $\infty$.  This  will then be equivalent  to fitting the
synthetic data (generated with a theoretical upper limit) with a model
without such a limit.  In the maps in Fig.~\ref{FIG1} we have kept the
imposed cut-off $X_{high}$ below the theoretical cutoff $x_{max}$, but
note  that  for real  data  (e.g.   in  the following  examples)  this
theoretical cut-off will be unknown.

Fig.~\ref{FIG2}  shows  the behaviour  of  the  fitted exponent  $\hat
\gamma$  as a function  of the  lower cut-off  $X_{low}$ for  the same
synthetic data samples as  in Fig.~\ref{FIG1}.  We have compared three
different ML estimations corresponding to three choices of $X_{high}$:

\begin{enumerate} 
\item  The  green circles  correspond  to  the  ML estimation  of  the
exponent  obtained   by  fixing   $X_{high}=x_{max}  =  10^8$   ,  the
theoretical upper limit. This is  nothing more than the profile of the
ML  exponent  map  along  the  vertical  right  border  at  $10^8$  in
Fig.~\ref{FIG1}.  This would  be the  correct  way to  perform the  ML
estimation if one  could have \textit{a priori} knowledge  of the true
upper cut-off.
\item  The  blue  triangles  (higher  symbols) correspond  to  the  ML
estimation obtained by  neglecting the existence of an  upper limit in
the  power-law  distribution,  i.e.  fixing  $X_{high}=\infty$.   This
would    precisely   correspond    to   the    method    proposed   by
Ref.~\onlinecite{Clauset2009}.
\item The red squares (lower  symbols) correspond to the ML estimation
obtained by fixing the higher cut-off of the data to the maximum value
found in the sample $X_{high}=X_{max}$.  It corresponds to the profile
of the map along the vertical black lines in Fig.~\ref{FIG1}.

\end{enumerate}

For small $N$ Method 3 underestimates the exponent because it neglects
the  fact  that  no  data  has been  observed  between  $X_{max}$  and
$x_{max}$. However,  for $N\gtrsim  10^4$, one can  see that  Method 3
renders exponents  that are very  similar to the correct  ones (Method
1).  For $N=10^6$  Method 3  is clearly  better than  Method  2, which
neglects the  existence of an  upper boundary to the  distribution: it
can be  seen that the  two coinciding estimation methods  (red squares
and green  circles) exhibit a larger  plateau (by more  than 1 decade)
than the blue triangles.

It is  also interesting to discuss  the strong fluctuations  of the ML
exponent close to  the diagonal of the maps, which  can be observed in
Fig.~\ref{FIG1}. These are due to the small size $n$ of the restricted
sample set that increases the statistical fluctuations.  We can locate
the region  where the standard deviation  of the ML  estimate is lower
than $\pm  0.05$ \cite{Note1} by  using the formulas  (\ref{error}) or
(\ref{errorclauset})  .   These low-error  regions  correspond to  the
areas below the black- and  red-curved lines, respectively.  As can be
seen,  both estimations of  the error  differ, even  in the  region of
small events.  The estimation that takes into account the existence of
both a low  and a high cut-off obtained  from Eq.(\ref{error}) gives a
better separation  between the regions with  meandering contour levels
and the smooth, flat plateau.

To conclude this  section, let us summarize what  we have learned from
the  analysis  of synthetic  data:  it does  not  make  much sense  to
increase the higher cut-off $X_{high}$  above the maximum value in the
data sample  $X_{max}$ unless we  have independent information  of the
theoretical limit $x_{max}$.  Therefore,  in the maps presented in the
following sections  we will  scan the higher  and lower  cut-offs only
within $X_{min}$ and $X_{max}$. Thus, the vertical right border of the
maps  will   coincide  with  the   vertical  black  line   plotted  in
Fig.~\ref{FIG1}.   In  addition,  we  will use  the  error  estimation
proposed by equation  (\ref{error}) and plot, on the  maps, the curved
line separating  the region  with error bars  greater than  $\pm 0.05$
(above the line) from the region with lower error bars (below).

\section{Seismological catalogue analysis}
\label{Seis}

The Gutenberg-Richter  law\cite{Gutenberg1944,Utsu2002} describing the
statistical distribution  of earthquake magnitudes is one  of the most
famous examples of a  scale-free phenomenon already discussed using ML
methods  in previous  works\cite{Kagan1991}. Theoretical  studies have
proposed      different     physical      models     (Burridge-Knopoff
\cite{Burridge1967},  Olami-Feder-Christensen \cite{Olami1992}, damage
rheology \cite{Ben-Zion2009},...)  and  framework theories such as the
so-called             Self            Organized            Criticality
\cite{Bak1987,Sornette1989,Main1996}   which  have  explained,   to  a
certain extent, the reasons behind this critical behaviour.  It is not
our purpose to  gain any understanding of seismology,  but only to use
some  of the  available earthquake  catalogues  in order  to test  the
behaviour of ML exponent maps.

Earthquakes  are  historically  characterized  by  a  quantity  called
magnitude $M$, which aims to  be a logarithmic measure of the ``size''
or ``energy  released'' during the  earthquake.  The Gutenberg-Richter
law\cite{Gutenberg1944} refers  to the number  of earthquakes $N_>(M)$
with a  magnitude larger  than a  certain value $M$  As a  function of
magnitude, the Gutenberg-Richter law can be written as:
\begin{equation}
N_>(M)\propto - b M
\label{Gutenberg}
\end{equation} 
with $b\sim 1.0$.
The  measurement of  earthquake  magnitudes and  energies  is still  a
challenging  issue for  seismology.  As  the medium  is  too large  to
collect a  significant amount  of radiated energy,  seismologists must
rely  on measurements  of the  sparse network  of seismic  stations in
order to locate  and estimate the ``size'' of  an earthquake.  Because
of this,  many different criteria  are used, depending on  the region,
earthquake energy  range, available instruments,  etc.  The catalogues
usually contain  mixed data corresponding to  different definitions of
the  magnitude  $M$.   These  definitions are  not  fully  equivalent,
especially for  small earthquakes\cite{Utsu2002,Scordilis2006}.  There
are also  different definitions of  the ``energy'' associated  with an
earthquake (all  of them are approximately  linearly related): seismic
moment, strain  energy drop, radiated  energy, etc..  In this  work we
will  use a  broadly accepted  formula\cite{Hanks1979} that  allows an
approximate  conversion of  the  different ``magnitudes''  $M$ to  the
minimum strain energy drop $E$ as:
\begin{equation}
\log_{10} E =1.5M + 4.8
\label{EdeM}
\end{equation}
where $E$  is the energy  in Joules.  Using the  Gutenberg-Richter law
(\ref{Gutenberg}) and Eq.~(\ref{EdeM}),  one can write the probability
density for earthquakes with energies between $E$ and $E+dE$ as:
\begin{equation}
p(E)dE \sim E^{-\epsilon} dE
\end{equation}
where  $\epsilon= 1+  (b/1.5) \simeq  1.67$ is  the  expected exponent
characterizing the power-law distribution of earthquake energies.

\begin{figure}[htb]
\begin{center}
\epsfig{file=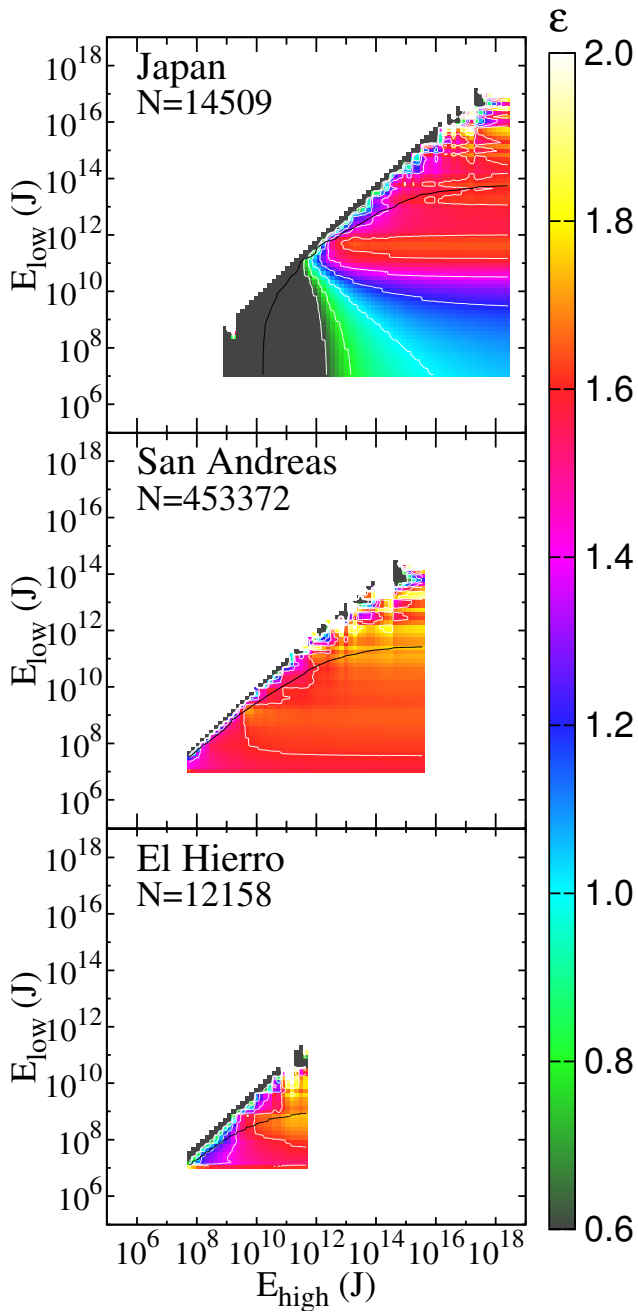,angle=270, width=8.6cm,clip=}
\end{center}
\caption{\label{FIG3} ML exponent  map corresponding to the earthquake
data from Japan,  the San Andreas fault and  El Hierro.  White contour
lines  are separated by  0.1 units.  The region  above the  black line
corresponds  to estimated  statistical  error bars  greater than  $\pm
0.05$.}
\end{figure}

\begin{figure}[htb]
\begin{center}
\epsfig{file=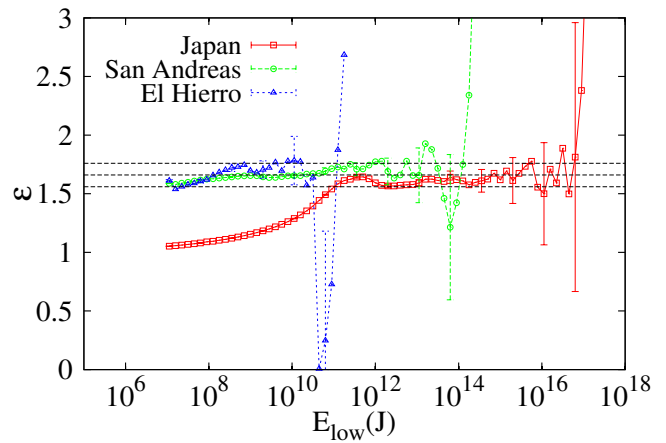, width=8.6cm,clip=}
\end{center}
\caption{\label{FIG4} Behaviour  of the ML  exponent as a  function of
the lower  cut-off $E_{low}$ for a  fixed higher cut-off  equal to the
maximum  value  in  the  sample  set.   Only  a  few  error  bars  are
indicated. The horizontal dashed lines show the theoretically expected
value 1.66 and an error bar $\pm 0.10$.}
\end{figure}

We  have  computed  the   ML  exponent  maps  corresponding  to  three
earthquake catalogues:
\begin{enumerate}
\item The subduction process taking place in the Japan Trench makes it
one of the most active seismological regions in the world. The area is
quite well documented because of the vicinity of the Japanese islands.
We studied  the exponent map corresponding to  the energy distribution
for  all the  seismological events  registered as  earthquakes  in the
ANSS\cite{ANSS}     catalogue      from     2000/01/01,00:00:00     to
2011/11/09,17:32:36  within  the  region  enclosed  between  latitudes
28$^{\circ}$N  and  48$^{\circ}$N  and longitudes  128$^{\circ}$E  and
148$^{\circ}$E.   The  registered  data  correspond to  the  $N=14509$
events  above  $M =  2.7$,  where  the  the T\={o}hoku  earthquake  of
2011/03/11 was the  most serious event with an  estimated magnitude of
$M=9.0$.
\item  San Andreas  fault system\cite{Okubo1987},  beneath  the region
occupied by the states of California and Nevada, there is probably the
most frequently  monitored seismic  region in the  world and  the best
documented  in catalogues.  We will  therefore  take it  as a  precise
example  of a  seismological strike-slip  process.  The  data analysed
here corresponds  to the seismic signals registered  as earthquakes in
the ANSS\cite{ANSS}  catalogue with its  epicentre within the  area of
latitudes  between  30$^{\circ}$N  and  42$^{\circ}$N  and  longitudes
between  114$^{\circ}$W and 126$^{\circ}$W  during the  period between
2000/01/01,00:00:00  and 2011/11/09,17:43:00.  In  order to  avoid the
presence  of  a possible  noise  background,  we  selected only  those
earthquakes  with a  magnitude  greater than  $M=0.4$.  The  strongest
earthquake  was  recorded on  2005/06/15  off  the  Coast of  Northern
California with  a magnitude of  $M=7.2$. The data set  has $N=453372$
events.
\item  As  a   completely  different  seismological  phenomenon  (very
localized  in space  and  time), we  considered  the recent  submarine
volcanic eruption of  La Restinga off the island  of El Hierro (Canary
Islands),  which  started  in   summer  2011.  The  volcanic  activity
triggered  an earthquake swarm\cite{Hainzl2003}  which is  expected to
have quite  different behaviour  from typical tectonic  processes.  We
considered  the data  obtained from  the IGN\cite{IGN}  catalogue from
2011/06/08, 00:52:00 until 2012/2/07,  12:00:00 in the region enclosed
by  latitudes from 26.8$^{\circ}$N  to 27.6$^{\circ}$N  and longitudes
from 17.85$^{\circ}$W  to 18.2$^{\circ}$W. The data  set has $N=12158$
events
\end{enumerate}
The three ML exponent maps are shown in Fig.~\ref{FIG3}.  We have kept
the same scales  in order to clearly reveal the  different size $N$ of
the statistical  samples. Fig.~\ref{FIG4}  shows the behaviour  of the
fitted exponent $\epsilon$ as a function of the lower cutoff $E_{low}$
when the  higher cut-off $E_{high}$ is  fixed to the  maximum value in
the sample set, i.e.  the profile  of the map along the vertical right
borders in Fig.~\ref{FIG3}.
%
%

The  first observation  is that  there  is an  almost perfect  plateau
exhibited by the  San Andreas data (middle diagram)  for a value close
to  the  expected  theoretical  value $\epsilon=1.67$.   Despite  some
deformation, indications of a plateau  are also observed for the other
two sets, Japan (top diagram) and El Hierro (bottom diagram).  For the
El  Hierro data  the coincidence  is remarkable,  given  the different
physical origins of the earthquake sequence.

A  second important  observation  is the  deformation  of the  plateau
(towards  low exponent values)  for the  Japan data  in the  region of
$E_{low}<10^{10}J$.  A  plausible explanation for  this deformation is
that the  statistics for small  earthquakes in the Japan  catalogue is
incomplete.  The  same tendency  can be observed  for the  San Andreas
data,  but for much  lower minimum  cut-offs $E_{low}<10^8  J$, almost
coinciding with the lower border of the map in Fig.~\ref{FIG3} (middle
diagram). The  oscillation of the  fitted exponent for the  Japan data
that  can be  seen on  the  maps (as  contour lines  with a  parabolic
horizontal shape starting from the right border) is also surprising as
well  as the  maximum  (about $E_{low}\sim  10^{11}-10^{12}J$) on  the
profile  shown  in  Fig.~\ref{FIG4}.   We  are unable  to  provide  an
explanation  for  this  behaviour,  but  it could  be  caused  by  the
different  methods   used  to  estimate   magnitudes  and/or  energies
depending on the earthquake magnitude range.

\section{Acoustic Emission Data Analysis}
\label{Exp}

As a  second set  of experimental examples,  we have focussed  on much
smaller   energy  scales.   Different   processes,  which   have  been
classified  as critical  or  crackling noise,  take  place in  solids,
exhibiting a  certain degree  of disorder when  driven by  an external
force    or    by    a    temperature    ramp.     Examples    include
superconductivity\cite{Wu1995},                               capillary
condensation\cite{Lilly1996},   acoustic    emission   in   structural
transitions\cite{Vives1994,Rosinberg2010,Planes2011}, Barkhausen noise
in magnetism\cite{Barkhausen1919,Durin2006}, fracture\cite{Rosti2009},
etc.   In  many  cases,  theoretical  studies  have  provided  general
frameworks      to     understand      the     origin      of     this
criticality\cite{Kardar1998,Fisher1998,
Sethna2001,Zaiser2006,Alava2006,Perez-Reche2008}.

The following three experimental  examples correspond to data recorded
using   the   Acoustic   Emission   (AE)   technique\cite{Scruby1987}.
Propagation of  cracks or the  sudden movement of  internal interfaces
generate acoustic waves in the ultrasonic range that propagate through
the solid and  that can be recorded by  appropriate transducers on the
surface.  The  AE method is equivalent  to the method  used to monitor
earthquakes,  but on  a  much  smaller scale.   It  is interesting  to
describe here some details in  order to understand the deviations that
will be observed in the maps.

The  most  common piezoelectric  transducers,  coupled  to the  sample
surface, generate a voltage signal $V(t)$ proportional to the speed of
the  incident elastic  wave  that is  amplified.   Using a  predefined
threshold   (above  the   unavoidable   experimental  electrical   and
mechanical  noise), it  is  possible to  define  individual AE  events
\cite{ManualPCI2}.  The beginning  of an event occurs at  a time $t_1$
when the voltage  exceeds the threshold.  The end  of the event occurs
when the signal  falls below the threshold at  $t_2$ and remains below
the threshold for more than  a certain pre-defined time called the Hit
Definition Time (HDT),  typically in the range of  10-100 $\mu$s.  The
fast  integration  of  the   $V^2(t)$  signal  from  $t_1$  to  $t_2$,
normalized  by a reference  resistance, renders  an estimation  of the
energy recorded by the transducer  which is assumed to be proportional
to the energy released by  the physical process generating the elastic
wave.

It is worth mentioning  some experimental limitations of most standard
set-ups:  (i) acquisition  systems,  due to  limited  memory, have  an
internal  maximum limit  on the  duration  of a  signal as  well as  a
maximum limit  on the  voltage that saturates  the amplifier.   In the
case that  such maxima are exceeded,  the signal is  truncated both in
voltage  and/or  duration.  This   represents  a  deformation  in  the
large-event region  and a  not totally sharp  cut-off in  the measured
energy  since   both  voltage   and  duration  can   be  independently
exceeded.  (ii)  A  second  experimental  problem  that  needs  to  be
considered,  is  due to  the  attenuation  of  ultrasounds inside  the
material  and the  distance from  the source  of the  AE event  to the
transducer.   If the studied  samples are  small (compared  to typical
length  scales  for exponential  attenuation  or  compared to  typical
transducer sizes) we expect that  data recorded by a single transducer
would not be very distorted by the distance to the source. However, if
samples   become  large,   the  quality   of  the   overall  power-law
distribution becomes  poorer.  One could then  use several transducers
to locate  the position  of the  source of the  event and  correct for
attenuation.     This    has    been    achieved   in    some    cases
\cite{Weiss2000,Vives2011}, thought,  in general, it  is a complicated
procedure.  The examples  below  correspond  to the  use  of a  single
transducer.   (iii)A  third problem  is  that  counting  of the  small
signals  is lower than  expected for  different reasons:  an important
fraction  of signals  that  happen to  be  very short  in time  and/or
amplitude cannot be  detected by the acquisition set-up  (they are too
short  for  the sampling  frequency  or  the  amplitude is  below  the
threshold value),  or some of the  small signals may  be overlapped by
the tails of previous signals due to dead-time HDT.

In the  cases analysed  in the following  three subsections,  the same
experimental  set-up  for the  acquisition  of  AE  has been  used:  a
PCI2-system from  the MISTRAS Group,  which consists of an  18-bit A/D
converter working at a base  sampling rate of 40 MHz.  The transducers
are also the same in the  three examples (micro80).  This makes the ML
exponent  maps  easy  to   compare.   The  amplification  factor,  the
threshold and  the number  of recorded signals  used in each  case are
different, given  the different noise conditions  and differing nature
of the studied phenomenon and driving force.  Although the acquisition
card is only 18 bits ($\sim  5$ decades), the measured energies by the
fast integration algorithm may theoretically extend many more decades.
In  order to  compare the  different studies,  we will  keep  the same
scales: from $10^{-17}$J to $10^{-9}$ J for energy and from 1.0 to 2.5
for the fitted exponent $\epsilon$.

\subsection{Mesoporous SiO$_2$ under compression}

The study of noise in  porous materials under compression is important
for   the  prediction   of  accidents   in  mining.   In   this  first
example\cite{Salje2011}  two  samples  of  a porous  material  SiO$_2$
(Vycor) with two parallel faces are compressed between two plates with
a lineally increasing load in time.   The AE sensor is attached to one
of the  plates.  There are two  sets of data  studied corresponding to
the  AE  events recorded  at  two loading  rates:  0.2  kPa/s and  1.6
kPa/s. The  study of  the influence of  the driving rate  is important
because, in  some cases,  it affects the  fitted exponents due  to the
overlap            of            large            and            small
avalanches\cite{Dahmen2003,Perez-Reche2004c}.   The   material  cracks
under  compression  and  the  recorded  AE signals  show  a  power-law
distribution of  energies.  In this case  a preamplifier of  60 dB was
used  and  the threshold  was  selected to  be  26dB.   The number  of
recorded signals was $N=11022$ and  $N=28652$ for the first and second
set, respectively.

\begin{figure}[htb]
\begin{center}
\hspace*{-10pt} \epsfig{file=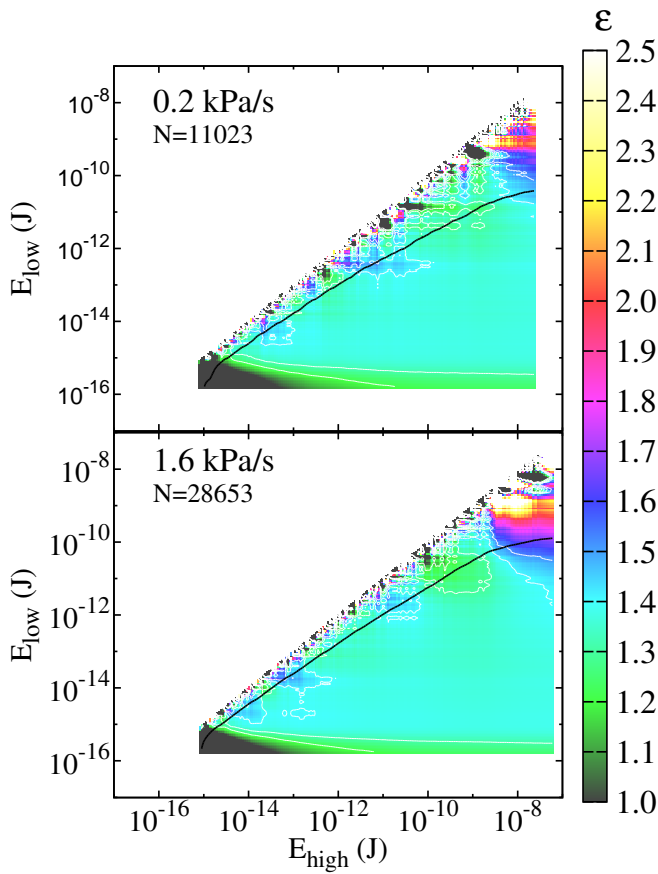,width=8.6cm,clip=}
\hspace*{10pt}
\end{center}
\caption{\label{FIG5}  ML  exponent   maps  corresponding  to  the  AE
recorded  during the  compression of  Vycor samples  at  two different
rates, 1.6kPa/s (above) and  0.2 kPa/s (below). White lines correspond
to the exponent  contour levels with a separation  of 0.1.  The region
above the black line indicates an estimated error bar grater than $\pm
0.05$.}
\end{figure}

Fig.~\ref{FIG5}  shows the  ML  exponent maps  obtained  with the  two
driving rates. It is clear that, in both cases, there is a vast region
of the map with a constant  plateau corresponding to a common value of
the  exponent close to  1.4. This  means that  the exponent  is robust
against  changes  in  the  cut-offs  by several  decades.   The  small
coloured spots  observed close to the  diagonal of the map  as well as
the large peak  in the upper right-hand corner  correspond to expected
statistical fluctuations, above the black line that indicates when the
statistical  error bar  becomes  larger than  $\pm  0.05$.  The  black
region in the bottom left-hand corner corresponds to the fact that the
exponent decreases when only an important fraction of very low signals
is included  in the analysis  which gives an erroneous  estimation for
the  reasons  explained above.   Basically,  no  other deviations  are
observed in this example.
%
\begin{figure}[htb]
\begin{center}
\hspace*{-10pt} \epsfig{file=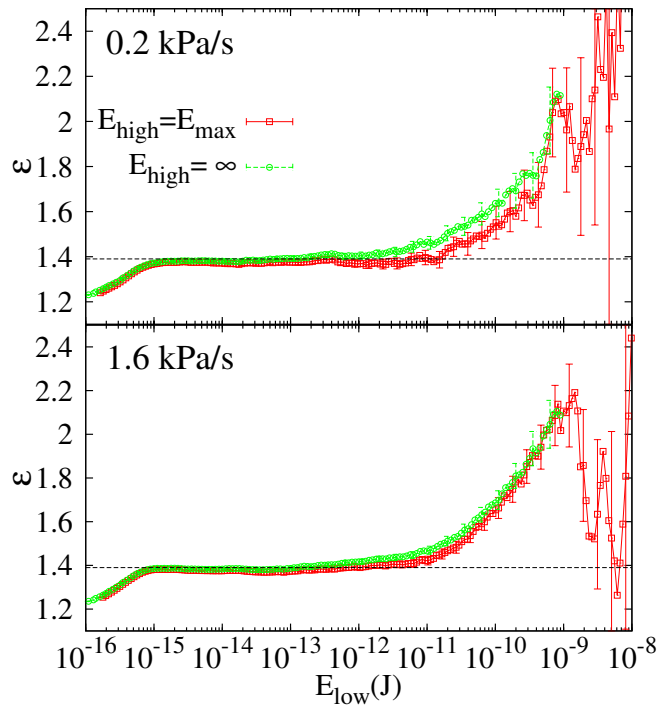, width=8.6cm,clip=}
\hspace*{10pt}
\end{center}
\caption{\label{FIG6} Profiles along the  right vertical border of the
exponent maps corresponding to  the Vycor experiments, compared to the
profiles  proposed   in  Ref.~\onlinecite{Clauset2009}.  Some  typical
estimated  error bars are  shown.  The  horizontal line  indicates the
value $\epsilon=1.39$ proposed in Ref.~\onlinecite{Salje2011}.}
\end{figure}

Fig.~\ref{FIG6} shows  the analysis of  the profiles of the  map along
the  right  vertical   border  ($E_{high}=E_{max}$)  compared  to  the
analysis proposed  in Ref.~\onlinecite{Clauset2009} assuming  no upper
limit for  the power-law distribution ($E_{high}=\infty$).   As can be
observed, the two  profiles are similar, but are  not identical.  They
show   a  clean   plateau  at   $1.39\pm  0.02$   for   three  decades
($10^{-15}-10^{-12}$).  The  profile obtained  by  fixing the  highest
cut-off to the maximum value  in the sample (i.e. along the right-hand
border  of  the ML  exponent  map  in  Fig.~\ref{FIG5}) shows  flatter
behaviour along, at least, one more decade.

Therefore,  one can  conclude that  this example  shows a  very robust
power-law  behaviour, comparable  to that  observed  for seismological
data.

\subsection{Structural transition in FePd}

In this second case we present an analysis of the AE recorded during a
martensitic  transition in  Fe-Pd \cite{Bonnot2008}.   This  alloy has
received  a lot  of theoretical  interest due  to the  simple symmetry
relation   between  the   high-temperature  phase   (cubic)   and  the
low-temperature phase  (tetragonal). The transition can  be induced by
applying an  external stress or by changing  temperature.  Two samples
were considered in the study:  a single crystal and a polycrystal with
the  same  composition  Fe$_{68.8}$Pd$_{31.2}$.   The  transition  was
induced by changing the temperature at different driving rates between
0.1K/min and 10K/min. Apart from the critical distribution of energies
during the  events, the AE  study revealed other  interesting features
(which  were  not  found  by  other techniques).  The  transition,  on
cooling, started at  the same temperature $M_s=246\pm2 K$  for the two
samples.   As soon  as the  transition  started the  formation of  new
tetragonal domains or the  advance of interfaces separating previously
nucleated  domains, generated  AE  events.  Due  to the  thermoelastic
behaviour\cite{Ortin1989} of  the transition the AE  extended for many
degrees  ($\sim 35  K$) until  the sample  was fully  transformed.  On
heating,  the reverse  process  occurred with  low thermal  hysteresis
($\sim 1$K).

In this case  AE was amplified $60$dB and the  threshold was chosen to
be 22dB.  The  four sets of data that we  will analyze here correspond
to a  driving rate  of $1K/min$  for both the  single crystal  and the
polycristal and  for cooling and  heating runs.  In order  to increase
statistics,  data were  accumulated over  20 ramps.  This accumulation
technique would correspond  to a real increase of  the statistics only
if the data recorded on each ramp is independent of the previous data.
This  assumption is  doubtful in  the case  of  structural transitions
since it has been demonstrated that samples exhibit a learning process
that  increases the  correlation  of the  signals between  consecutive
ramps \cite{Perez-Reche2004b}.   In this case  the ramps corresponding
to the  same driving  rate were not  strictly consecutive  since other
driving  rates were  added in  between, thus  the independence  of the
recorded data is not clear.

The total  number of recorded signals was  $N=171056$ (single crystal,
cooling), $N=111840$ (single crystal heating), $N=192596$ (polycrystal
cooling) and $N=58501$ (polycrystal heating). Typically, the number of
AE events is not symmetric  during heating and cooling ramps and there
is a lack of understanding of this phenomenon\cite{Vives1994}.

The  study \cite{Bonnot2008}  concluded  that in  the  four cases  the
energies  of the  individual AE  events were  power-law  distributed (
$\sim 4$ decades  for the single crystal and $\sim  3$ decades for the
polycrystal).  The exponents were fitted  using the ML method and were
almost  independent of the  heating/cooling rate.   The values  of the
cut-offs  were  selected  in  the  region where  the  statistics  were
sufficiently high.   For the single  crystal both cooling  and heating
ramps  exhibited  an exponent  compatible  with  a  value of  $1.64\pm
0.10$. For the polycrystal a clear deviation of the exponent was found
when the heating  ($2.0 \pm 0.1$) and cooling ramp  ($1.59 \pm 0.10 $)
were found.  So far, we have no explanation for this deviation.

\begin{figure}[htb]
\begin{center}
\hspace*{-10pt} \epsfig{file=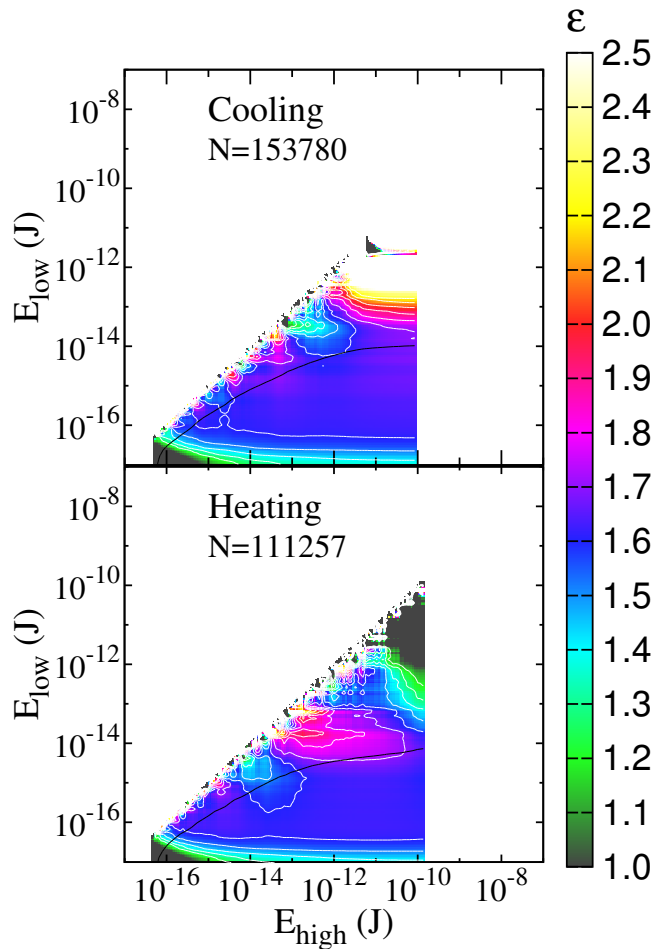,width=8.6cm,clip=}
\hspace*{10pt}
\end{center}
\caption{\label{FIG7} Exponent maps for  AE in the FePd single crystal
corresponding to heating ramps (below) and cooling ramps (above).  The
region  above  the black  line  corresponds  to  estimated error  bars
greater than $\pm  0.05$, after correcting by a  factor of $\sqrt{20}$
due to the possible correlations between measurements (see text).  The
contour lines (in white) are separated by 0.1 units.}
\end{figure}

\begin{figure}[htb]
\begin{center}
\hspace*{-10pt} \epsfig{file=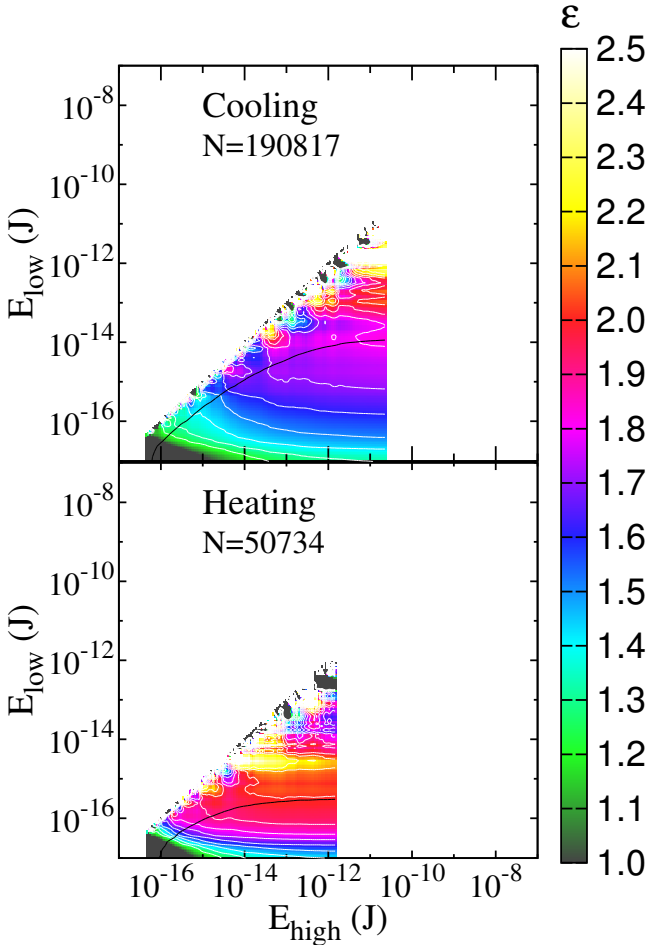,width=8.6cm,clip=}
\hspace*{10pt}
\end{center}
\caption{\label{FIG8}  Exponent maps  for AE  in the  FePd polycrystal
corresponding to heating ramps  (below) and cooling ramps (above). The
region above the black line  corresponds to estimated error bars above
0.05, after correcting by a  factor of $\sqrt{20}$ due to the possible
correlations between  measurements (see  text). The contour  lines (in
white) are separated by 0.1 units.}
\end{figure}

Fig.~\ref{FIG7}   and   ~\ref{FIG8}   show   the  ML   exponent   maps
corresponding  to the  four  cases.  The  graphs  above correspond  to
cooling data and the graphs below  to heating data. As can be observed
the  coloured spots  attached  to the  diagonal,  which correspond  to
statistical fluctuations  are larger (in  absolute terms) than  in the
previous example.  This suggests that, although the recorded number of
signals is $\sim 10$ times larger, most probably data corresponding to
the different 20 runs were correlated and did not effectively increase
the statistics.  Thus, the estimated  error bars were,  most probably,
underestimated by a factor of $\sqrt{20}$. Therefore, in order to plot
the curved black line on the  map separating the zone with large error
bars, we  have required that $\sqrt{20} \sigma_{\hat  \gamma} = 0.05$.
It can be  seen that, indeed, the curved line  separates the zone with
fluctuating contour lines  from the flat region. By  observing the map
corresponding  to the  heating runs  (Fig.~\ref{FIG8} bottom)  for the
polycrystalline sample one clearly sees that the large fitted exponent
$\sim 2.0$ is doubtful and most  probably is a consequence of the lack
of  statistics. The  region with  small enough  error bars  (below the
black line) is very small and it is difficult to identify any plateau.
For the  cooling ramps (Fig.~\ref{FIG8}  top) the plateau is  not very
clear but at least the separation of the contour lines is much wider.
%

\begin{figure}[htb]
\begin{center}
\epsfig{file=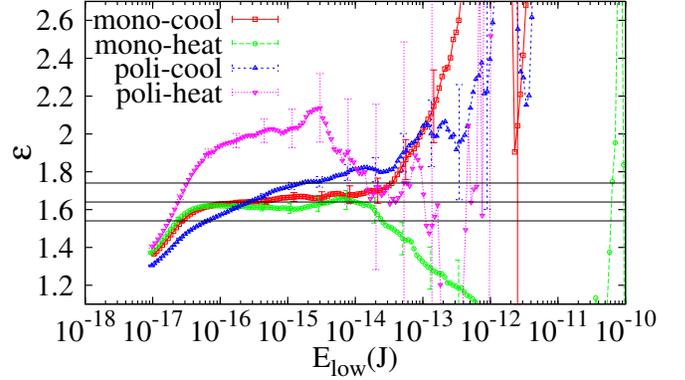, width=8.6cm,clip=}
\end{center}
\caption{\label{FIG9} Exponent  values versus lower  cut-off $E_{low}$
  for AE in  FePd. The horizontal lines show  the value $1.64\pm 0.10$
  proposed in Ref.~\cite{Bonnot2008}}
\end{figure}

Fig.~\ref{FIG9}  shows  the  corresponding  profiles along  the  right
vertical border of the map. Error bars have been increased by a factor
of $\sqrt{20}$  compared to  the ones reported  in the  original paper
\cite{Bonnot2008}.   The dashed lines  indicate the  proposed exponent
$1.64\pm0.1$. The hypothesis that this might be a common value for the
exponent corresponding  to the four cases  cannot be ruled  out due to
the  lack of  statistics.  More  measurements for  the polycrystalline
sample would be required to fully clarify this point.

\subsection{Structural transition in CuZnAl} 

This  third example corresponds  to a  recent study  of AE  during the
martensitic    transition    in    a   CuZnAl    shape-memory    alloy
\cite{Gallardo2010}.  In  this case  the transition is  also thermally
induced from  a monoclinic multivariant structure  at low temperatures
to a cubic structure at high temperatures. The purpose of the analysis
performed  in Ref.  \onlinecite{Gallardo2010}  was not  to demonstrate
the  power-law distribution  of the  AE events  (which had  been shown
previously in different studies \cite{Vives1995,Carrillo1998}), but to
compare the  observed exponent with the exponent  corresponding to the
very large  avalanches that  can also be  recorded by  a sophisticated
calorimetric study.   The sample  studied was, therefore,  larger than
those  studied  previously.  This,  as  explained  above  may lead  to
distortion of  the power law.  Furthermore, a larger  sample typically
requires a more powerful  temperature control set-up, involving higher
electric currents  and thus involving greater noise.   Since the study
was focussed on large avalanches and noise was high, the amplification
was set  to a much lower  factor, 40 dB,  and the threshold to  45 dB.
The data set analyzed consists of $N=17936$ signals corresponding to a
unique heating ramp.

\begin{figure}[htb]
\begin{center}
\epsfig{file=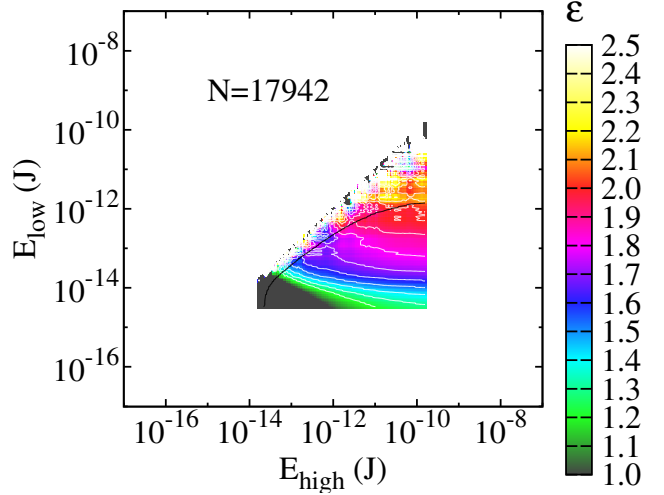,width=8.6cm,clip=}
\end{center}
\caption{\label{FIG10}   ML   Exponent   map  corresponding   to   the
martensitic transition in a CuZnAl sample during a heating ramp.}
\end{figure}
\begin{figure}[htb]
\begin{center}
\epsfig{file=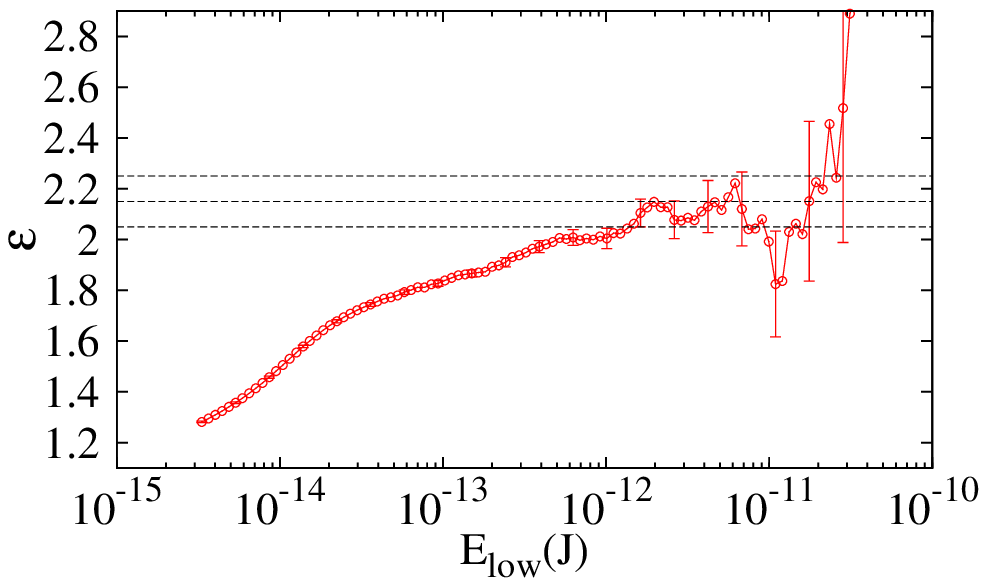, width=8.6cm,clip=}
\end{center}
\caption{\label{FIG11} ML  fitted exponent as a function  of the lower
cutoff  $E_{low}$  when considering  a  higher  cut-off  equal to  the
maximum  recorded signal  in the  data sample.   The  horizontal lines
indicate the value $2.15\pm0.10$.}
\end{figure}

As can  be seen in Fig.~\ref{FIG10},  the exponent map  shows a poorer
quality   without   any  clear   plateau.    The   profile  shown   in
Fig.~\ref{FIG11}, corresponding  to the  right vertical border  of the
map, suggests  an incipient  plateau around $\epsilon=2$  for slightly
more than  one decade,  but this is  not fully conclusive.   Thus, the
exponent map analysis in this case indicates that either the number of
signals is too small to draw conclusions about the power-law behaviour
of the  data or the distribution  shows a deformation  probably due to
the fact  that the sample was  too large and  attenuation introduces a
length scale.  Note,  however, that we are not  fully compromising the
results pointed out Ref.  \onlinecite{Gallardo2010} .  The conclusions
were based not only on AE  signals, but also on the coincidence of the
observed incipient  plateau at $\epsilon=2$ with  the plateau observed
by a different experimental technique.

\section{Simulation data}
\label{Sim}

As  a  last case  we  will  analyze  data corresponding  to  numerical
simulation of  a lattice model.   This is an  interesting illustrative
case of  the advantages of ML  maps compared to  the previous proposed
analysis  assuming no  upper limit.  This is  because  the distortions
affecting  the  large avalanche  region  are  not  due to  measurement
problems,  but are intrinsic  to the  finite size  of the  model.  The
study  corresponds  to  the  3D  Gaussian  Random  Field  Ising  Model
(3D-GRFIM) driven by an external field $H(t)$ with metastable dynamics
at  zero temperature  \cite{Sethna1993}.  The model  is  based on  the
original Ising model with the addition of random internal fields $h_i$
acting on  each spin $S_i=\pm 1$.   The values $h_i$  are quenched and
distributed   according   to   a   Gaussian  probability   density   $
\rho(h)dh=\frac{1}{\sqrt{2\pi}  R} e^{\frac{-h^2}{2R^2}}dh$  with zero
mean and variance $R^2$.  The  parameter $R$ is usually referred to as
the amount  of disorder in the  system. The Hamiltonian  of the system
reads:
\begin{equation}
\mathcal{H}=- \sum_{\langle  i,j\rangle} S_i  S_j - \sum_i^N  h_i S_i-
H(t) \sum_i^N S_i
\end{equation}
where the $S_i$ spin variables  are defined on a regular cubic lattice
and  the first  sum  extends over  all  nearest-neighbour pairs.   The
simulations of the model start from a saturated configuration $\lbrace
S_i  \rbrace=-1$ and  $H=-\infty$ .  The field  is  then adiabatically
increased and the spins flip according to the local relaxation rule:
\begin{equation}
S_i = sign \left ( \sum_j S_j + h_i + H(t) \right )
\end{equation}
where  the sum  extends  over all  the  $z=6$ neighbours  of the  spin
$S_i$. With  this metastable dynamics  the system evolves  following a
sequence  of  magnetization jumps  (avalanches)  occurring at  certain
fixed values of the external  field separated by periods of inactivity
in which the field is increased without producing any spin flip.

The model has  been widely used as a prototype model  for the study of
avalanche dynamics.   It has  been successful in  explaining different
features of the magnetization process in ferromagnets: the presence of
rate independent hysteresis, the  return point memory property and the
existence  of Barkhausen noise  \cite{Sethna1993}.  Extensions  of the
model  have been  also used  for the  understanding of  other athermal
first-order phase transitions.\cite{Vives2001,Cerruti2008}

Here we will  focus our attention on the  distribution $D(s,R)$ of the
sizes $s$ (number  of flipped spins) of the  avalanches obtained along
the magnetization  process from  $H=-\infty$ to $H=+\infty$,  i.e. the
so-called  integrated distribution.   In the  thermodynamic  limit and
when the amount of disorder $R$  is tuned to the critical value $R_c$,
this     distribution     is     expected     to    be     a     power
law\cite{Sethna1993,Perkovic1999,Perez-Reche2003,Perez-Reche2004}
characterized by  a critical exponent  called $\tau'=\tau+\sigma \beta
\delta$
\begin{equation}
D(s,R_c) = \frac{ s^{-\tau'}}{\sum_{s=1}^{\infty} s^{-\tau'}}
\end{equation}
For  values  of   disorder  above  $R_c$  it  is   expected  that  the
distribution of  avalanche sizes is exponentially damped  and thus, in
the  thermodynamic  limit, the  discontinuities  in the  magnetization
$\Delta  m= s/L^3$  vanish.   Below  $R_c$, it  is  expected that  the
distribution  of avalanches  is also  exponentially damped,  but there
will  exist a  unique massive  avalanche with  a size  proportional to
$L^3$ that is responsible  for a magnetization discontinuity as should
occur given the first-order character of the phase transition.

In the  numerical simulations on  a finite lattice ($L\times  L \times
L$) with periodic  boundary conditions, nevertheless, the distribution
of  avalanche  sizes $D(s,R,L)$  behaves  quite differently.   Several
effects  deform  the power-law  character  at the  ``pseudo''-critical
point:

On the one hand, avalanche sizes  are limited from above by the finite
size  $s<L^3$.   The fact  that  in  all  critical phenomena  with  an
associated diverging correlation  length, distortions occur well below
the  limit $L^3$ is  well known.   Among other  reasons, close  to the
critical  point  $R_c$, the  avalanches  are  expected  to be  fractal
\cite{Perkovic1995, Perez-Reche2004} and  thus exceed the lattice side
$L$ when  its size is much  smaller than $L^3$.   Such avalanches that
expand  the lattice  side $L$  in, at  least, one  dimension,  are the
so-called  spanning avalanches.   For  a cubic  lattice with  periodic
boundary  conditions, they  can be  easily classified  as 1D-spanning,
2D-spanning or 3D-spanning depending  on whether they span the lattice
side in 1,2 or 3 spatial dimensions.

On  the other hand,  some small  avalanches (sometimes  called lattice
animals within  percolation theory \cite{Stauffer1994})  may occur for
probabilistic reasons even very far away from the critical point.  For
any value  of $R$ and  $H$ it is  possible to compute  the probability
that  two  neighbouring  spins  (surrounded by  negative  spins)  flip
simultaneously.   The same  computation  can be  carried  out for  the
probability of a group of three, four, etc.  neighbouring spins with a
certain topological configuration to flip simultaneously.  It is clear
that in such a computation  the number of configurations in which spin
clusters (lattice animals) with a  certain size $s$ can be constructed
plays  a  fundamental role.   Such  a  phenomenon  has (a  priori)  no
connection  with  the  critical  point.   The  total  number  of  such
non-critical avalanches is, nevertheless very large (it increases with
$L^3$)  and  renders  a  distribution  of small  avalanches  that  has
general, fast  decreasing behaviour  with $s$, but  is non-monotonous.
At the critical point this effect overlaps with the proposed power-law
distribution and thus shows a  deformation in the small-size region of
the  distribution $D(s;R,L)$,  which typically  can be  observed below
$s\sim 10$.  Our  goal here is to show how  these distortions at large
and small sizes show up on the ML exponent map.

Before the discussion  of our results, it is  important to recall that
the  exact  values of  $R_c$  and  $\tau'$  have not  been  definitely
established.  The  problems are precisely due  to the fact  that it is
difficult to  deal with  spanning avalanches in  the simulations  on a
finite lattice. There have been two main approaches:

\begin{enumerate}

\item Dahmen and coworkers found  $R_c=2.16 \pm 0.03$ and $\tau'= 2.03
\pm  0.03$   by  performing  a  scaling  collapse   of  the  avalanche
distribution, neglecting the fact that there should be a dependence on
$L$ in  the distribution.  This was first  done \cite{Perkovic1999} on
very  large  systems (up  to  $1000^3$)  but  averages over  very  few
realizations  of   disorder.  The  collapses   clearly  revealed  that
corrections  to  scaling  were  needed.   Later\cite{Liu2009}  similar
collapses were  done which included  a unique system size  $64^3$.  In
both  cases,   the  data  were  restricted  to   amounts  of  disorder
$2.25<R<6.0$  well above  $R_c$.   By this  method  they avoided  most
spanning avalanches in  the simulations but paid the  price of working
too far  from the critical  point and so  they had to  extrapolate the
value for the exponent.

\item Furthermore,  there have been studies precisely  focussed on the
behaviour of  spanning avalanches  which analyse how  they concentrate
close  to the  critical  point\cite{Perez-Reche2003, Perez-Reche2004}.
Such studies have performed finite-size scaling analysis of the number
of  spanning  avalanches and  have  obtained  a  higher value  of  the
critical amount  of disorder  $R_c=2.21$.  To do  so, they  proposed a
method  (called  method-2  in Ref.   \onlinecite{Perez-Reche2004})  to
separate  the  3D-spanning  avalanches  that will  correspond  to  the
massive avalanches that clearly  disturb the power-law distribution of
avalanche sizes in  the region of large events,  close to the critical
point.  Such  3D-spanning avalanches are identified when  they are the
unique spanning  avalanches in  the full-field excursion.  This allows
them to be  filtered from the statistical analysis.  The reason behind
the separation method is that when massive avalanches occur, they fill
such a  large fraction of  the system that  they do not allow  for any
other  spanning avalanche  to  take  place.  It  was  shown that  this
filtering method,  although not perfect, gave  more consistent results
in the finite-size scaling analysis  than the method of discarding all
the spanning avalanches.

\end{enumerate}

The differences between the two  above approaches are even more subtle
and difficult  to summarize here.  Let  us simply remark  that the two
analysis  use a  slightly different  scaling variable  to  measure the
distance   to    the   critical   point   and    that   the   exponent
$\tau'=2.03\pm0.03$  proposed   in  Ref.~\onlinecite{Perkovic1999}  is
interpreted  not as  a true  critical  exponent, but  as an  effective
exponent $\tau'_{eff}$ in Ref.~\onlinecite{Perez-Reche2003}.  Instead,
they  propose that  close to  the critical  point the  distribution of
avalanche sizes (neglecting the  massive avalanches) will be dominated
by non-critical  lattice animals  (in the small  s region) and  by the
so-called  non-spanning critical avalanches  (in the  intermediate and
large s  region). Only  the last kind  of avalanches may  exhibit true
power-law behaviour with an exponent $\tau_{nsc}=1.65$ at $R_c=2.21$.

We have performed numerical simulations with system sizes ranging from
$L=32$   to   $L=256$   and   values   of   $R$   within   the   range
$2.15<R<2.24$. For every  size $L$ and every $R$,  averages were taken
over 2000  configurations of  the random fields  for $L \leq  128$ and
over 1400 for $L=256$. We  used the RANECU random number generator and
a Box Muller-Polar Marsaglia algorithm to generate the Gaussian random
fields.   Simulations of the  metastable dynamics  were done  with the
sorted  list  algorithm \cite{Kuntz1999}.  The  typical  sizes of  the
sample sets of recorded avalanches (corresponding to each single value
of $R$) range from $10^6$ for $L=32$ to $10^9$ for $L=256$.

\begin{figure}[htb]
\begin{center}
\hspace*{-10pt} \epsfig{file=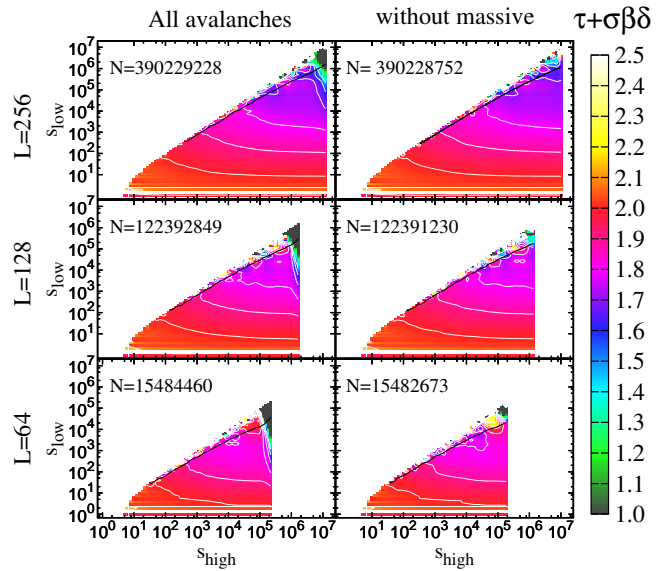,width=8.6cm,clip=}
\hspace*{10pt}
\end{center}
\caption{\label{FIG12} ML exponent  maps corresponding to the 3D-GRFIM
for $R=2.20$ and increasing system sizes from $L=64$(below) to $L=256$
above.  The left-hand  column shows  the maps  that where  obtained by
considering  all  the  recorded  avalanches,  whereas  the  right-hand
column, the  method proposed in  Ref.~\onlinecite{Perez-Reche2004} was
used  to  suppress the  massive,  large  non-critical avalanches.  The
region above  the black line (very  close to the diagonal  of the map)
corresponds to an estimated error bar of $\pm 0.05$.  Contour lines in
white are separated by 0.1 units}
\end{figure}
\begin{figure}[htb]
\begin{center}
\epsfig{file=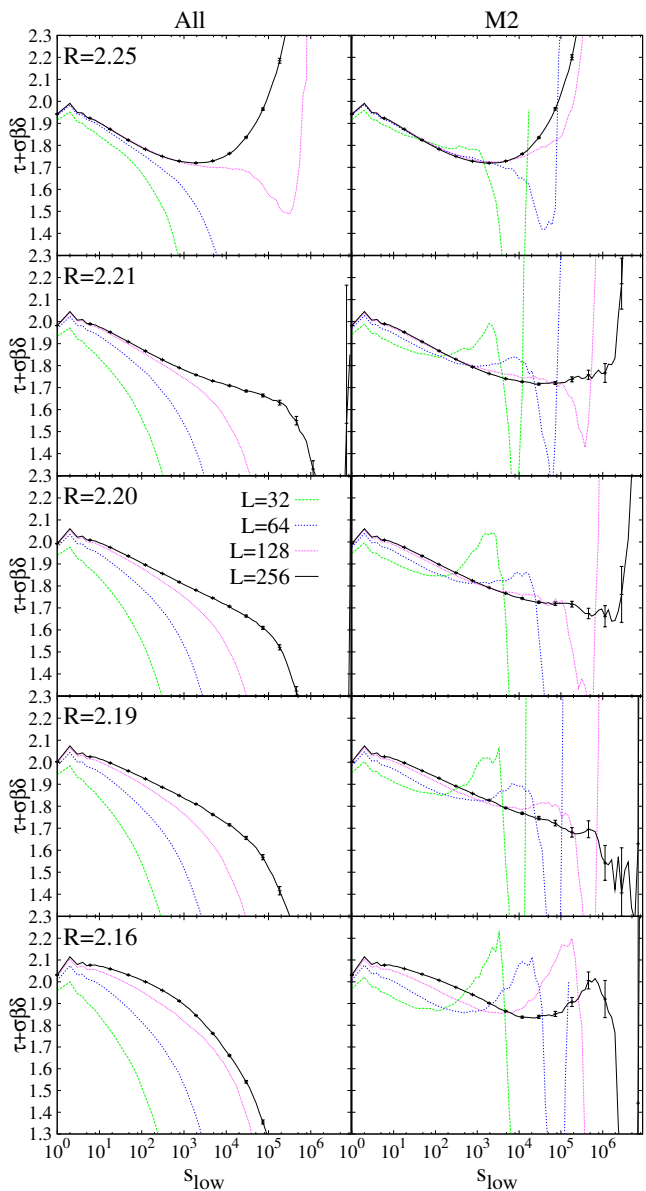,width=8.6cm,clip=}
\end{center}
\caption{\label{FIG13}  Profiles of  the  ML exponent  maps along  the
vertical right  border for different values  of $R$ and  $L$.  Data in
the left  column corresponds  to the full  set of avalanches.   On the
right column  data corresponds  to the same  sets after  filtering the
massive    avalanches    following     the    method    proposed    in
Ref.~\onlinecite{Perez-Reche2004}.}
\end{figure}

Fig.~\ref{FIG12} shows the exponent maps corresponding to $R=2.20$ and
increasing values of $L$. The first column corresponds to the exponent
obtained by  considering all the  avalanches and the second  column to
the  analysis after  the  suppression of  the  massive avalanches,  as
explained in Ref. \onlinecite{Perez-Reche2004}.  As can be observed in
the first  column the maps  display an approximately  triangular black
region  in the  upper right-hand  corner that  also extends  along the
right edge  which is due  to including of such  ``massive'' avalanches
that  they  strongly  distort  the  possible  power-law  distribution.
Massive avalanches overpopulate the large $s$ region and thus decrease
the value of the fitted  exponent.  Such a deformation disappears when
the  data  is  filtered,  as  can  be seen  in  the  right  column  in
Fig.~\ref{FIG12}.

Close  to  the  bottom  horizontal  boundary  of  the  maps,  exponent
oscillations (seen as a sequence of parallel horizontal contour lines)
can also be observed in all the graphs.  These oscillations are due to
non-critical avalanches (lattice animals) which are always present and
difficult to subtract.

Besides these two observed deformation regions, it is not clear that a
clean plateau  exists close to  the bottom right-hand corner,  with an
area  increasing  with $L$.  The  maps show  a  slow  decrease of  the
exponent  from $2.1$ towards  $1.7$ when  increasing the  lower cutoff
$s_{low}$, and this effect does not seem to disappear for large system
sizes.   Instead, a  region  with a  value  of the  exponent close  to
$1.6-1.7$ seems to  develop close to the upper  corner when the system
size is increased up to $L=256$.  This effect is clearer when the data
has been  filtered (right column) and  suggests that a  true power law
may only  be observed  when small avalanches  are not included  in the
analysis.

Fig.~\ref{FIG13}  shows the profiles  of the  maps obtained  by fixing
$s_{high}=s_{max}$,  for different  values of  $R$ between  $2.16$ and
$2.25$ and $L$ from $32$ to  $256$. The column on the left corresponds
to the analysis  of all the avalanches and the column  on the right to
the  analysis after filtering  the massive  avalanches.  Again,  it is
difficult to  find any clear evidence  of a plateau  growing with $L$.
The  only exception seems  to be  for $R=2.20-2.21$  in the  region of
intermediate and large cut-offs ($10^4<s_{low}<10^6$), where a plateau
seems to form and become  broader when $L$ increases. This plateau has
a height approaching $\tau'\simeq 1.7$. It can be observed not only on
the filtered data  (right column), but also as  an inflection when the
avalanches are not filtered (left column).

The profiles  in Fig.~\ref{FIG13} also  allow the oscillations  in the
small avalanche region ($s_{low}<20$) to be observed. The peaks in the
exponent  correspond   to  even  sizes.  This   indicates  that  small
avalanches  with odd  sizes occur  with  a higher  frequency than  the
frequency corresponding to a perfect power law.

Note  also  that  for   the  critical  value  $R_c=2.16$  proposed  in
Ref.~\onlinecite{Perkovic1999},  the  possible  plateau  at  a  height
$\tau'\sim 2.03$  does not exhibit  a clear tendency to  increase with
$L$: it is  very much distorted by odd-even  fluctuations and, already
for $s_{low}=10^2$,  has clearly decreased for all  the studied system
sizes.

In  our opinion,  a  final  understanding of  the  behaviour of  these
distributions can  only be achieved  after a full  finite-size scaling
analysis   of  the  distribution   $D(s;R,L)$  (involving   the  three
variables), which is beyond the  scope of this paper. Nevertheless, in
view  of the  above observations  it  seems that  the distribution  of
avalanche sizes at the critical point shows two contributions: (i) the
actual power-law behaviour  corresponding to the non-spanning critical
avalanches  with  $\tau'\simeq  1.7$  that  can only  be  observed  at
intermediate and large avalanche sizes, and (ii) the contribution from
non-critical avalanches  (lattice animals),  in the small  $s$ region,
which   distorts   the    exponent   towards   the   effective   value
$\tau'_{eff}\simeq  2.03$.   This scenario  is  not incompatible  with
previous numerical studies.

\section{Summary and conclusions}
\label{Con}

\begin{figure}[htb]
\begin{center}
\epsfig{file=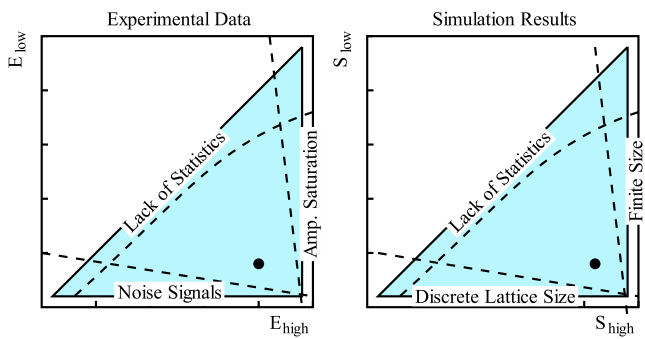,width=8.6cm,clip=}
\end{center}
\caption{\label{FIG14}  Schematic representation  of ML  exponent maps
for experimental  data (left) and  simulation data (right).  The lines
indicate  the  regions  where   we  can  expect  deformations  of  the
theoretical plateau. The dot indicates the best values of the cut-offs
for the estimation of the exponent.}
\end{figure}

The aim  of this  paper has  been to illustrate  the usefulness  of ML
exponent maps in order to  study distortions of power-law behaviour to
the critical  distributions of events.  Such  distortions are expected
to occur  for most experimental  and numerical simulation data  due to
different reasons.  Fig.~\ref{FIG14}  shows a schematic representation
of the  main conclusions of  this paper.  On  the left we show  the ML
exponent  map that  one can  expect for  experimental data.   When the
sample set is small, a  lack of statistics creates deformations of the
theoretical plateau close  to the diagonal axis of  the map, which can
be  bounded by  a proper  estimation  of the  statistical error  bars.
Noise  and undercounting  of small-size  events renders  a deformation
region  starting in  the  bottom left-hand  corner  and which  extends
horizontally along  the bottom  border.  Saturation of  the amplifiers
and counters also  deforms the plateau in the  upper right-hand corner
and along the right edge.  Between these three boundaries there should
be a region with a  well-defined plateau.  The black dot indicates the
best values of  the high and low cut-off for  the determination of the
critical exponent.  It should  be mentioned that,  if the size  of the
sample  set  is  too small,  such  an  ideal  situation might  not  be
achieved.

The right  plot shows  a similar scheme  for numerical  simulations of
lattice systems.  The deformation  regions are similar to the previous
ones.  Along the  bottom edge  we find  the footprint  of  the lattice
character of  the model that deforms the  perfect power-law behaviour.
Along  the vertical right  edge we  find deformations  associated with
finite-size effects, which in some cases can be partially corrected.

The examples analyzed in  this work have included: three seismological
catalogues corresponding to Japan, the San Andreas fault and El Hierro
volcanic activity, three sets  of Acoustic Emission data corresponding
to  the  fracture  of  Vycor  under  compression,  a  cubic-tetragonal
structural  transition  in  FePd  and  a  cubic-monoclinic  structural
transition  in CuZnAl and,  finally, numerical  simulations of  the 3D
Random Field  Ising model. These three  case studies of  the maps have
allowed the good  quality of the data to be  checked or hypothesis for
the observed  distortions to be  proposed. The studies have  also been
used  to   suggest  improvements  to   measurements  and/or  numerical
simulations.

\begin{acknowledgments}
The computations presented  in this work have been  carried out on the
IberGRID infrastructure of the  Spanish network, e-Ciencia.  This work
has received financial support from the Spanish Ministry of Innovation
and Science (project MAT2010-15114).  We acknowledge fruitful comments
from Antoni Planes and Pol Lloveras.
\end{acknowledgments}

\end{document}